\documentclass{aa}  
\bibpunct{(}{)}{;}{a}{}{,} 
\usepackage{graphicx,xcolor}
\usepackage{txfonts}
\usepackage[version=4]{mhchem}
\usepackage{subcaption}

\begin{document} 

   \title{Infrared photoprocessing of porous amorphous solid water with a free electron laser}

   \author{J.G.M. Schrauwen\inst{1}
   \and S. Ioppolo\inst{2}
   \and L. Slumstrup\inst{2}
   \and J. D. Thrower\inst{2}
   \and B. Redlich\inst{1,3}
   \and H.M. Cuppen\inst{4}
          }

   \institute{HFML-FELIX laboratory, IMM, Radboud University,
              Toernooiveld 7, 6525 ED Nijmegen, The Netherlands
        \and
            Center for Interstellar Catalysis, Department of Physics and Astronomy, Aarhus University, 8000 Aarhus C, Denmark
        \and
            Deutsches Elektronen-Synchrotron (DESY), Photon Science, Notkestr. 85, 22607 Hamburg, Germany
        \and
            Institute of Molecules and Materials (IMM), Radboud University, Heyendaalseweg 135, 6525 AJ Nijmegen, The Netherlands\\
            \email{herma.cuppen@ru.nl}
             }

   \date{Received ...; accepted ...}

 
 \abstract
   {Water ice is the most abundant ice in the interstellar medium (ISM). 
   Previous studies on porous amorphous solid water (pASW) have shown that the infrared radiation field in the ISM modeled with infrared free electron laser (FEL) irradiation can result in changes in the ice towards a more ordered structure, characterized by a change in the OH-stretching vibration band.}
   {This paper aims to study the restructuring in more detail by varying the irradiation intensity and irradiation time for irradiations on-resonance with the OH-stretching, bending and libration modes.
   In parallel to the restructuring, the desorption of \ce{H2O} is monitored during the irradiations.}
   {The research presented in this paper considers 42 infrared FEL irradiations on pASW samples deposited at two different rates, using the infrared FELs of the HFML-FELIX laboratory.
   Restructuring is studied using reflection absorption infrared (RAIR) spectroscopy, with a mass spectrometer monitoring the desorption of \ce{H2O} in an ultra-high vacuum system.}
   {Desorption and restructuring occur in parallel, with the strongest effects observed for irradiation on-resonance with the OH-stretching vibration. 
   Both processes exhibit two distinct regimes: an initial strong restructuring and desorption, characterized by an exponential decrease in the intensity of the desorption peaks within the first 5~s of irradiation, followed by a second regime of less efficient, yet continued restructuring and constant desorption in an extended area, involving energy dissipation through the hydrogen-bonding network.
   The trends in restructuring and desorption, depending on the absorbed energy during irradiation, are significantly different, suggesting separate processes.}
   {The second regime of constant desorption and restructuring could be significantly more relevant for processes in the ISM compared to the first regime.
   Future infrared irradiation experiments should therefore focus more on this second regime.}


   \maketitle
%

\section{Introduction}
In the late 1930s, \citet{eddington1937} proposed the presence of solid water in the interstellar medium (ISM), assuming that the high hydrogen content of the ISM would favor the formation of molecules such as \ce{H2O} from non-ionized atoms. 
This prediction was confirmed decades later in 1973, with the detection of absorption features associated with the OH-stretching vibration \citep{gillett1973}.
Following this initial observation, water was discovered to be the most abundant solid-phase molecule in space, with its distinctive OH stretch feature observed across a wide variety of astrophysical environments \citep{boogert2015}.

Water-ice formation is expected to already start in diffuse interstellar clouds \citep{cuppen2007,lamberts2014,cazaux2010}.
In such environments, low temperatures allow atomic oxygen and hydrogen to adsorb onto the surfaces of interstellar dust grains. 
Subsequent atom-addition reactions on these grain surfaces form \ce{H2O} molecules \citep{tielens1982,cazaux2010}.
The newly-formed \ce{H2O} is readily dissociated by the interstellar ultraviolet (UV) radiation field in diffuse clouds and only a small steady-state population of ice likely persists under these conditions \citep{cuppen2007,lamberts2014}, which has now also been confirmed by JWST observations \citep{zeegers2025}.
As the cloud becomes more dense, \ce{H2O} ice is shielded from the UV field, allowing \ce{H2O} ice to accumulate in thicker layers on dust grains. Irradiation in the range between 91 an 191 nm, which can trigger photodissociation and photodesorption, reduces to $9.4\times10^{-2}$ photons$\cdot$s$^{-1}\cdot$cm$^{-2}$  for an extinction of 10~mag \citep{whittet2003,mathis1990} and the cosmic-ray-induced photon flux takes over with $3-15\times10^{3}$ photons$\cdot$s$^{-1}\cdot$cm$^{-2}$  \citep{cecchi1992}. Both fluxes are considerably lower than the unattenuated flux of $1.7\times10^{8}$ photons$\cdot$s$^{-1}\cdot$cm$^{-2}$ \citep{mathis1983}.

However, radiation in the infrared wavelength regime is not as strongly attenuated and the photon flux in the infrared region exceeds the UV flux by several orders of magnitude ($1.0\times10^{10}$ photons$\cdot$s$^{-1}\cdot$cm$^{-2}$ between 2.5 and 20 $\mu$m) \citep{mathis1983}. Even when accounting for the wavelength-dependent absorption cross section, the flux of absorbed photons is significant with $3.4\times10^{7}$ absorbed photons$\cdot$s$^{-1}\cdot$cm$^{-2}$ or $8.9 \times 10^{-13}$ J$\cdot$s$^{-1}\cdot$cm$^{-2}$.
Infrared-induced photoprocessing of water ice is therefore an important process to consider, as it may represent a significant pathway for water-ice photodesorption in cold regions of molecular clouds and beyond the \ce{H2O} snow line in protoplanetary disks and it may process the remaining ice, leading to a changed spectroscopy and reactivity of the ice mantle.
The structure of the ice mantle determines the available binding sites on the ice surface for other species to adsorb to or diffuse over. The hydrogen bonding structure of water ice further stabilizes reaction intermediates and products by absorbing excess reaction energy. 
This stabilization is facilitated by the rapid energy dissipation through the hydrogen bonding network. The efficiency of this process depends on the structure of the network, which is different for crystalline and amorphous solid water \citep{post2015,sudera2020,cuppen2022}.

In our previous studies, we investigated vibrational energy dissipation in amorphous solid water (pure and mixed) using on-resonance infrared free electron laser (FEL) irradiation \citep{noble2020,cuppen2022,schrauwen2024,schrauwen2025,schrauwen2025CH4}. 
In all studies, we consistently observed a characteristic change in the 3~$\mu$m OH-stretching band profile, attributed to a reorganization of \ce{H2O} molecules into configurations with, on average, more hydrogen bonds per molecule. 
Resonant excitation of the OH-stretching mode can also induce photodesorption of water in crystalline \ce{H2O} \citep{krasnapoler1998,focsa2003,noble2020} and of CO on top of amorphous solid water (ASW) \citep{slumstrup2025}. 

\citet{krasnapoler1998} irradiated multilayer crystalline water ices with picosecond infrared FEL pulses, to develop new surface and depth-profiling techniques for photoablation therapy. 
They observed strong resonant desorption, with the maximum yield occurring at 3.0~$\mu$m. 
A single 3~$\mu$s FEL macropulse of 1~mJ desorbed roughly half of a 1~$\mu$m-thick ice layer in the first irradiation pulse at 110~K.
In this paper, we extend these studies to amorphous water ice with the aim of qualitatively explaining its photodesorption behavior and describing the restructuring in the ice in more detail. 
In addition to on-resonance excitation of the OH-stretching vibration, we also investigated resonant excitation of the bending and libration modes to explore the mode dependence of the photodesorption and restructuring processes.

\section{Experimental methods}
\begin{figure}
    \centering
    \includegraphics[width=\linewidth]{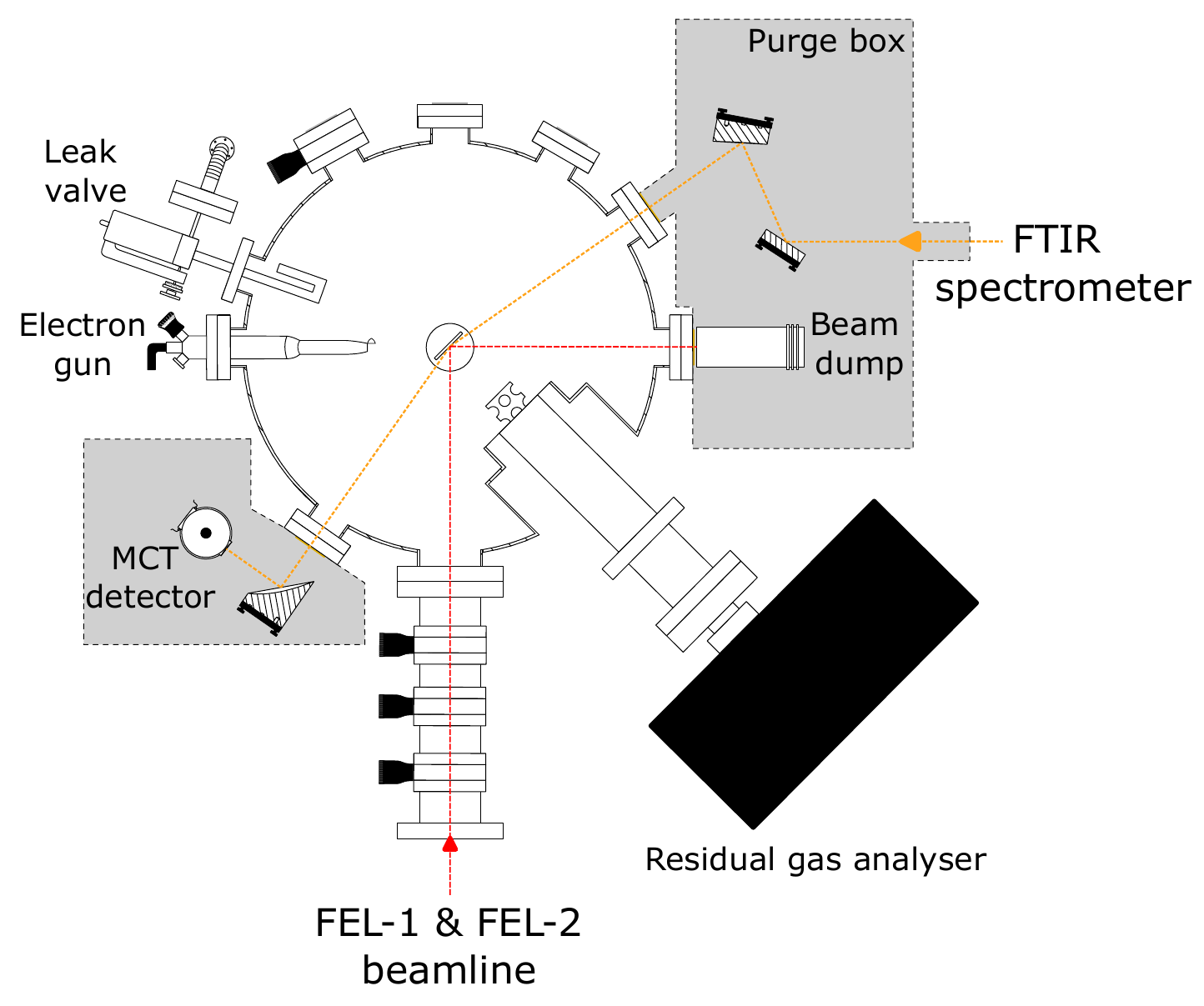}
    \caption{Schematic top view of the Laboratory Ice Surface Astrophysics (LISA) chamber connected to the FELIX beamline in US-1 at HFML-FELIX.}
    \label{fig:LISA}
\end{figure}

Experiments were performed with the Laboratory Ice Surface Astrophysics (LISA) chamber, a user station at the HFML-FELIX laboratory in Nijmegen, The Netherlands.
LISA is an ultra-high vacuum system (base pressure $1\cdot10^{-9}$~mbar at room temperature) designed for studying energetic processing of interstellar ices coupled to the beamlines of free electron laser (FEL) 1 and 2 that supply monochromatic and high-intensity infrared light over a broad wavelength range. 
A schematic overview of the LISA chamber is shown in Figure \ref{fig:LISA}.
The LISA chamber is discussed in more detail in \citet{ioppolo2022} and \citet{schrauwen2024} and in this section, we list only the details of the experiments reported here.

Porous amorphous solid water (pASW) samples are deposited from the gas-phase (from liquid; deionized and purified through multiple freeze-pump-thaw cycles) on the gold-coated copper substrate kept at the lowest achievable temperature of the closed-cycle helium cryostat ($\sim$9.4~K) in the center of the LISA chamber through an all-metal leak valve and a bent dosing line to ensure homogeneous background deposition.
Two pASW ices were studied with different deposition rates; a slow deposition of approximately 3 hours at a constant pressure of $1\cdot10^{-7}$~mbar and a fast deposition of roughly 4 minutes at $5\cdot10^{-6}$~mbar. The resulting spectra are given in Appendix \ref{app:spec_comp}.

Using that 1 Langmuir corresponds to a 1~s deposition at $1\cdot10^{6}$ Torr, the depositions resulted in a roughly 900~L deposition for the slow-deposited ice and a roughly 1050~L deposition for the fast-deposited ice.
From the current data set, the thickness of the ices cannot be determined accurately. 
However, an estimation based on the general assumption for \ce{H2O} at cold temperatures that 1~L=1 monolayer=3~\AA{} suggests that the layers are about 270$\pm30$~nm thick.
This is thick enough to exclude any surface effects from the gold-coated substrate.

The longer deposition time to achieve the same thickness can have two opposing effects: the longer relaxation time before a molecule covers another could promote a more ordered or compact structure, while the longer dissipation time of the adsorption energy could result in less heating during deposition, leading to reduced reordering of the ice with incoming molecules \citep{Speedy:1996,Smith:1997A,Hudson:2026}. 
Therefore, a difference in structure between the two samples with different deposition conditions could be anticipated.
In practice, the ices were spectroscopically very similar, as shown in Appendix Figure \ref{fig:slow_fast_TPD}, and also the irradiation had very similar effects. 
Still, a slight shift of 6~cm$^{-1}$ to lower wavenumbers in the OH-stretching vibration is observed for the slow deposition.
We expect this to indicate that the slow deposited ice is more stable compared to the fast deposited ice, since this shift is consistent with spectral shifts observed during mild annealing and is attributed to a more preferred hydrogen bonding configuration \citep{noble2020,Smit:2017B}.

In the analysis, we used the results of the irradiations on the fast and slow-deposited ices as complementary measurements, indicating the expected spread in the measurements based on difference in the deposition conditions for spectroscopically similar ices.
During all depositions, the growth of the ice was monitored by Fourier-transform reflection absorption infrared (RAIR) spectroscopy at a RAIR angle of 10$^\circ$ and the gas phase composition was tracked by a residual gas analyzer (RGA; Hiden HAL3F PIC) mass spectrometer operating in multiple-ion detection (MID) mode.

After deposition, the pASW samples were left to stabilize until no differences were observed in the RAIR spectra taken 5 minutes apart.
Subsequently, the ice was irradiated with FEL-2 -- a pulsed laser that produced macropulses of $\sim6$~$\mu$s at 5~Hz composed of micropulses at a 1 GHz repetition rate for these experiments. 
Using FEL-2, irradiations were performed on resonance with OH-stretching, bending, and libration vibrations of the pASW samples at 3~$\mu$m, 6~$\mu$m, and 12~$\mu$m, respectively.
As sketched in Figure \ref{fig:LISA}, the p-polarized FEL-2 beam impinged on the substrate with a 45$^\circ$ angle, resulting in an estimated irradiated spot of 0.6~mm$^2$, 2.4~mm$^2$, and 4.0~mm$^2$ for irradiations at 3~$\mu$m, 6~$\mu$m, and 12~$\mu$m, respectively, as reported in \citet{slumstrup2025}.

Different irradiation experiments were performed at these wavelengths, varying the irradiation time (between 10 s and 5 minutes) and the FEL intensity (full energy and attenuation with 3 dB, 5 dB, and 13 dB resulting in a reduction of the FEL intensity of 50~\%, 70~\%, and 95~\%, respectively).
The intensity of the FEL pulse depends on the irradiation wavelength and was between 40 and 135~mJ, with generally lower intensities at the shorter wavelengths. 
A vertical translation stage allowed for ten independent, non-overlapping measurement spots per prepared ice sample. 
For these experiments, some irradiations were performed at the same spot to study potential saturation of the irradiation effects.

The effect of infrared FEL irradiation was investigated by taking RAIR spectra just before and just after irradiation, in the range of 5000-500~cm$^{-1}$ with a resolution of 0.5~cm$^{-1}$ and a total of 256 co-added scans. 
The difference of the two spectra, baseline-corrected with a linear interpolation of selected baseline points (post-irradiation minus pre-irradiation; hereafter denoted as difference spectra), indicated the changes in the shapes of the vibrational bands, with a positive signal denoting an increase in population at that wavelength and a negative signal denoting a loss of population. 

Here, it should be noted that the area probed by the infrared spectrometer is significantly larger than the FEL beam.
The probed area of the infrared spectrometer was 90~mm$^2$, which is 20 times larger compared to the largest FEL spot at 12.0~$\mu$m. 
The fraction of exposed ice material with respect to the probed material is wavelength dependent, resulting in a higher sensitivity for longer wavelengths.The difference in area is corrected by dividing the relevant quantities by the size of the irradiation spot so that all changes are given in terms of column densities. The difference in sensitivity is reflected in the different error bars per data point.
The effect of this difference is clearly visible in Figure~\ref{fig:desIRQMS} where the strongest spectral difference is observed for 12.0~$\mu$m, while the corrected effect per irradiated area is rather low, as will be discussed in Section~\ref{sec:fit} and Figure~\ref{fig:time_Ac}.

In addition to the RAIR spectra, the effect of the FEL irradiation was also studied by monitoring the gas phase during the irradiation using the mass spectrometer operating in MID mode, tracking only $m/z$~18 as a function of time with a dwell time of 5~ms and a settle time of 0~ms. 
These settings resulted in a sampling frequency of \ce{H2O} in the gas phase of 195~Hz, recording 38 data points between the FEL macropulses.
Despite a settle time of 0~ms, the mass spectrometer still exhibited a $\sim$0.2~ms interval per data point during which no measurements were taken, due to the data processing and loading by the MASsoft software (Hiden Analytical). 

Although this could potentially lead to a loss of data, a comparison with the recently developed 
multichannel scaler (MCS) mode at the LISA chamber that synchronizes the mass spectrometer with the FEL frequency
and does not have the uncontrolled downtime, reveals that the two regimes in the MID data, which we will discuss elaborately in this paper, are not an artifact of the absence of synchronization with the FEL and the downtime in the MID mode.
This new mode for the mass spectrometer became available only after the measurements for this study were performed and is not yet fully optimized for desorption measurements. Figure \ref{fig:MCSvsMID} in Appendix \ref{app:MCS} presents in more detail the results of the MID and MCS modes of the mass spectrometer and shows that they are relatively similar.
As such, all analysis in this paper is based on mass spectrometry measurements in MID mode.

\section{Results and discussion}
\subsection{On-resonance irradiation of three vibrational modes}
\begin{figure*}
    \sidecaption
    \includegraphics[width=12cm]{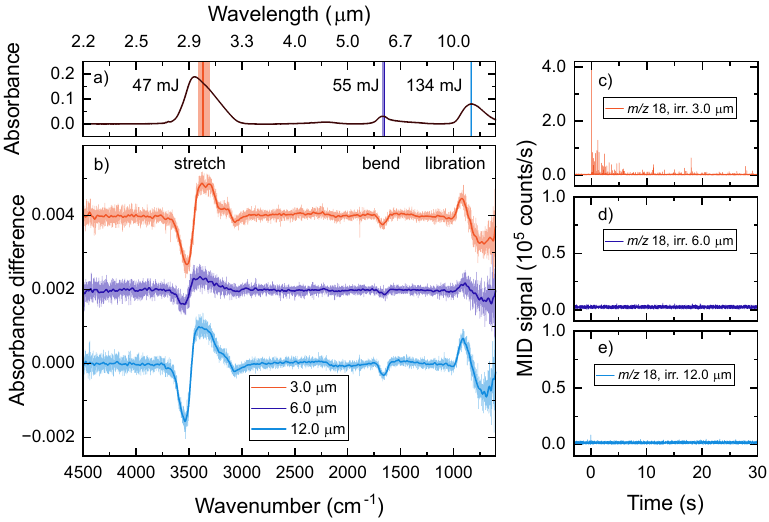}
    \caption{Effects of a 30 s irradiation on-resonance with the OH-stretching vibration (3.0~$\mu$m, orange), the bending vibration (6.0~$\mu$m, blue) and the libration mode (12.0~$\mu$m, cyan) of porous amorphous solid water b) as recorded by the RAIR difference spectra and c-e) as recorded by the mass spectrometer measuring \textit{m/z}~18 during the irradiation.
    A pre-irradiation spectrum of pASW is shown in a) with the overlap of the FEL beam with the specific vibrational mode, as indicated by the colored bars, with the dark vertical line denoting the central frequency.
    The difference spectra in b) are offset for clarity, and a smoothed version of the difference spectra is shown in a darker shade in front of the raw data.}
    \label{fig:desIRQMS}
\end{figure*}

The pASW samples were irradiated on-resonance with the OH-stretching, bending and libration modes of \ce{H2O} by FELIX FEL-2. 
Figure \ref{fig:desIRQMS} shows three exemplary irradiations of equal time duration (30~s) on the three different modes and the effect of the irradiation as recorded in the infrared spectra and by the mass spectrometer.
The data corresponding to irradiation on-resonance with the OH-stretching (3.0~$\mu$m), bending (6.0~$\mu$m), and libration (12.0~$\mu$m) modes are displayed in orange, blue, and cyan, respectively.
This color coding for irradiations on-resonance with the three different vibrational modes is used throughout this paper.

Figure \ref{fig:desIRQMS}a) shows the overlap of the FEL with the specific vibrational mode as vertical colored bars with the central frequency of the FEL shown as a darker solid line.
Because of the FEL characteristics, irradiation at longer wavelengths results in a beam with a narrower frequency bandwidth.
The average macropulse energy during the respective irradiations is shown next to the central frequency line. 
In addition to the three exemplary irradiations in Figure \ref{fig:desIRQMS}, an additional 39 irradiations were performed to study the effect of irradiation intensity and duration in detail.

Figure \ref{fig:desIRQMS}b) shows the infrared difference spectra corresponding to the three on-resonance irradiations.
The raw difference spectra are overplotted with their FFT filter smoothing curves in a darker color to aid interpretation.
All three irradiations show the characteristic down-up profile observed for the irradiation of pASW before in the OH-stretching region \citep{noble2020,cuppen2022}.
This profile corresponds to the reorganization of the \ce{H2O} molecules to a configuration with, on average, more hydrogen bonds per molecule. The OH-stretch region is highly sensitive to the local hydrogen bonding environment of the water molecules and specific frequencies have been attributed to specific hydrogen bonding patterns \cite{noble2020,Smit:2017B}. Here, the spectral change is consistent with change from a triply H-bonded to a quadruply H-bonded environment.
This reorganization suggests a more `crystalline' environment, but since infrared spectroscopy only probes the local environment, crystallization cannot be confirmed here.
Molecular dynamics simulations of a model pASW system did not find any changes in the long-range order after irradiation \citep{cuppen2022}. 
Hence, crystallization of the system is unlikely, yet a restructuring takes place that increases the local order.

The down-up feature in the OH-stretch profile (spectrum at 3000-3700~cm$^{-1}$) is clearly weaker for irradiation at the bending mode (6.0~$\mu$m, purple curve) compared to irradiation of the stretching and libration modes (red and blue curves). 
This can be partly explained by the difference in absorption coefficient for the different modes, which is substantially lower for the bending mode. 
Based on the extinction coefficient in \citet{rocha2023}, typical absorption coefficients are $1.6\cdot10^4$~cm$^{-1}$, $0.1\cdot10^4$~cm$^{-1}$ and $0.4\cdot10^4$~cm$^{-1}$ for irradiation at 3.0~$\mu$m, 6.0~$\mu$m and 12.0~$\mu$m, respectively.
As such, the bending vibration has a four times smaller absorption coefficient compared to the libration mode, and more than 10 times smaller compared to the stretching vibration, rationalizing the weak changes observed in Figure \ref{fig:desIRQMS} for irradiation at 6.0~$\mu$m. 
Moreover, at longer irradiation wavelengths, the irradiation spot of the FEL is larger. 
At 12.0~$\mu$m the irradiation influences an area that is almost ten times that of the irradiation of 3.0~$\mu$m. We will correct for the size of the irradiated area throughout this paper by dividing the relevant quantities, such as integrated areas and fitting coefficients, by the irradiated area.
The infrared difference spectra are not scaled.

Still, the OH-stretch region in the difference spectrum shows less substructure upon irradiation at the bending vibration compared to irradiation at the libration mode, and even less than at the stretching vibration.
The ratio between `up' and `down' also varies between the 3.0~and 12~$\mu$m irradiation, with a roughly equal down-component, but a larger up component for the 12~$\mu$m irradiation.
This suggests that more \ce{H2O} has desorbed at 3.0~$\mu$m compared to 12~$\mu$m irradiation.
We will further analyze these spectral differences in the next section by fitting the difference spectra.

Besides the restructuring, the MID traces shown in Figure \ref{fig:desIRQMS}c-e) reveal desorption of $m/z$~18 (\ce{H2O}) during the irradiation.
Clear desorption is only observed for irradiation at 3.0~$\mu$m in Figure \ref{fig:desIRQMS}c). 
For irradiation at 12.0~$\mu$m (Figure \ref{fig:desIRQMS}e), only a minimal desorption signal is observed just at the start of the irradiation. 
This is substantially lower than at 3.0~$\mu$m, which corresponds with the less prominent negative part compared to the positive part of the restructuring in the OH-stretching region for the 12~$\mu$m irradiation compared to the 3.0~$\mu$m irradiation in Figure \ref{fig:desIRQMS}b).
No desorption is observed for the 6.0~$\mu$m irradiation.
To the best of our knowledge, this is the first report of photodesorption of \ce{H2O} by on-resonance infrared irradiation for a pure pASW ice, although hints of \ce{H2O} desorption were observed previously for a low-temperature layered \ce{H2O}-CO system \citep{slumstrup2025}.

Every spike in the MID trace recorded during the irradiation at 3.0~$\mu$m corresponds to a FEL macropulse interacting with the pASW ice.
The first macropulse results in the largest number of desorbing \ce{H2O} molecules, and after that the intensity of the spikes decreases more or less exponentially.
Most of the desorption occurs within the first 10 seconds of the irradiation, but desorption spikes of lower intensity are still observable until the end of the irradiation.
This exponential decrease in the intensity of the desorption spikes over time 
has also been observed in other ices studied with an infrared FEL, such as CO in or on \ce{H2O} and \ce{H2O}:\ce{CH4} mixtures \citep{ingman2023,slumstrup2025,schrauwen2025CH4}.

The data suggest two different desorption regimes: one that follows an exponential decrease during the first 5~s ($\sim$25 macropulses) and a second that shows more or less constant continued desorption until the end of the irradiation.
For desorption of a multilayer of ice, one would expect a constant desorption signal with every macropulse, indicative of a zeroth-order process.
This zeroth-order process will persist as long as the ice stays in the multilayer regime.
However, the exponential decrease hints at desorption governed by a first-order process. 
This suggests that the population of water molecules available for desorption decreases with continued irradiation. 
This can either be because the ice leaves the multilayer regime or because only loosely-bound surface species can desorb, which means that only a limited population is available for desorption. 
\citet{krasnapoler1998} observed the desorption of the entire ice, clearly leaving the multilayer regime, but this is unlikely in our experiments, as restructuring is observed along with desorption, which should not be observable when all material has desorbed. 
Moreover, after the initial strong desorption, a continued, constant desorption signal is observed, indicating that material is still available for desorption.

\begin{figure}
    \centering
    \includegraphics[width=\linewidth]{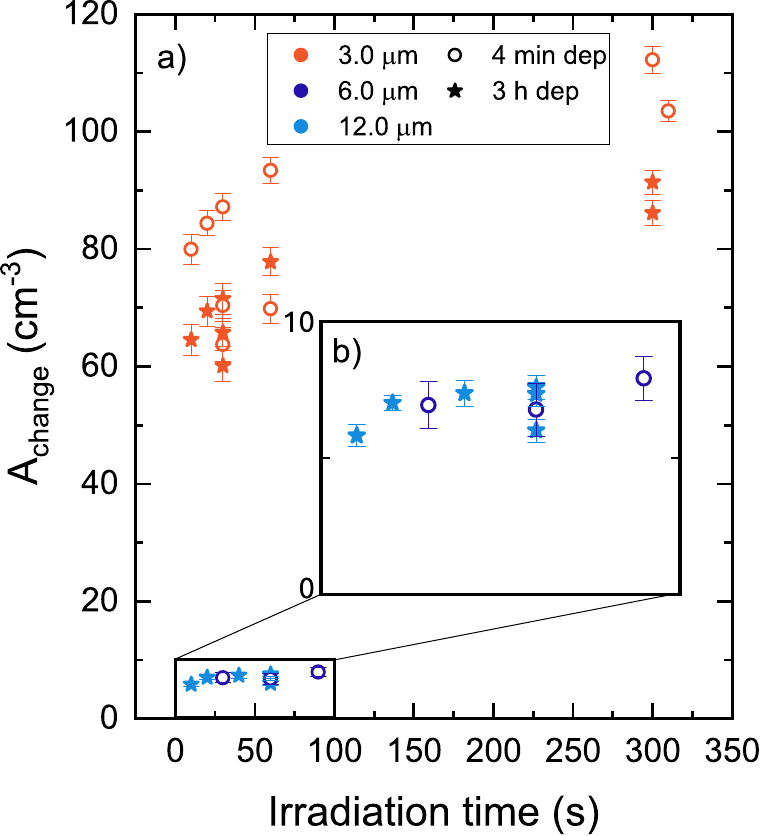}
    \caption{Integrated areas in wavenumber (cm$^{-1}$) of the region of the change in the OH-stretch region in the infrared difference spectra for the average FEL intensity irradiations of $\sim47$~mJ for irradiations at 3.0~$\mu$m (orange), $\sim114$~mJ for irradiations at 6.0~$\mu$m (blue) and $\sim114$~mJ for irradiations at 12.0~$\mu$m (cyan) corrected for the irradiated area (cm$^{-2}$). As such the unit of $A_\text{change}$ is cm$^{-3}$ (wavenumber per cm$^2$). The integrated areas are presented as a function of the time duration of the FEL irradiation. Irradiations performed on the pASW ice from a slow deposition (3 h) are presented as solid stars, and irradiations performed on the pASW ice from a fast deposition (4 min) are presented as open circles. The error bars are determined from the difference between the integrated area obtained from the raw difference spectra and from difference spectra smoothed with an FFT filter. b) shows a zoom-in on the 6.0~$\mu$m and 12.0~$\mu$m irradiations.}
    \label{fig:time_Ac}
\end{figure}

The two regimes observed in the mass spectrometry data are also observed in the changes in the infrared difference spectra.
Figure \ref{fig:time_Ac} shows a selection of irradiation experiments performed for different irradiation durations, but with similar macropulse energies per irradiation wavelength.
The extent of the restructuring is parametrized by the absolute spectral area of change ($A_\text{change}$) observed in the OH-stretching vibration in the difference spectra, which is divided by the irradiated area on the ice surface to obtain the extent of the restructuring per cm$^{2}$. 
This correction is important as Figure~\ref{fig:desIRQMS} shows very similar areas of change for the 3.0~$\mu$m and 12.0~$\mu$m irradiations, but, since the irradiated area is roughly 10 times larger for the 12.0~$\mu$m irradiation compared to the 3.0~$\mu$m experiments, the area of change per cm$^{2}$ in Figure~\ref{fig:time_Ac} is roughly 10 times lower for the 12.0~$\mu$m irradiation experiments.
With this correction, the changed area for the 12.0~$\mu$m irradiations becomes comparable to changes at 6.0~$\mu$m.

In Figure \ref{fig:time_Ac}, the $A_\text{change}$ obtained for irradiations on the fast-deposited ice (4-minute deposition) and the slow-deposited ice (3-hour deposition) are shown with different symbols.
It seems that the fast-deposited ice has consistently higher $A_\text{change}$ values compared to the slow-deposited ice, but both clearly follow the same trend for irradiations at 3.0~$\mu$m.
The higher $A_\text{change}$ values are likely not only the result of the slightly thicker ice that is obtained for the fast deposition, as the difference in $A_\text{change}$ of a maximum of 30~\% is significantly larger compared to the maximum 10~\% difference in thickness.
Therefore, we postulate that the slow-deposited ice is more compact and results in weaker changes, while the restructuring process remains the same, as shown by the down-up profile and the similar trends compared to the fast-deposited ice in Figure \ref{fig:time_Ac}.

As such, we observe that the compaction of the ice in the case of the slow-deposited ice (indicated in all Figures with stars) only dampens the effect of the irradiation.
This is not unexpected considering that the down-up profile we observe is the same type of peak shift as observed in the spectrum of the more compact ice.
Likely, part of the population available for restructuring in the ice is already ordered in the more compact ice, reducing the total effect of the infrared irradiation.
As the compaction of the ice only seems to result in weaker restructuring, we will not separate the two different ice depositions in our analysis.
Instead, they present the error margin for nearly identical pASW ices with a more or less compact structure.

Figure \ref{fig:time_Ac} reveals that immediately after 10 s of irradiation, the absolute area of change is at $\sim$70~\% of its maximum value, at all wavelengths. This corresponds to the first regime observed in the mass spectrometry data, where most desorption occurs.
After this first regime, a gradual increase in the absolute area of change is observed, which matches the second regime of slow changes seen in the mass spectrometry data.
Although these regimes observed in both the mass spectrometry data and the infrared difference spectra do not necessarily have the same origin, the processes of restructuring and desorption are likely closely related. The decrease in desorption efficiency may therefore be partly related to the simultaneous local restructuring of the upper ice layers, as previous electron-irradiation studies have shown that more ordered water-ice structures can respond less efficiently to energetic processing \citep{Zheng:2007,Mifsud:2022}, although the resonant IR excitation used here is mechanistically distinct from such radiolytic processes.
We will investigate this further in Section \ref{sec:trends}.

\subsection{Fitting of the difference spectra}
\label{sec:fit}
We devised a fitting routine for our infrared difference spectra to better understand the observed down-up profile and the underlying process of restructuring. 
This is based on the assumption that the restructuring can be described as global heating of the ice. 
We therefore recorded a small database of reference spectra by performing a temperature-programmed desorption (TPD) experiment on an unirradiated pASW ice deposited during 4 minutes at $5\cdot10^{-6}$~mbar.
Although the desorption was monitored during the TPD, our analysis focuses on the infrared spectra recorded during the temperature ramp.
The TPD infrared spectra were measured at temperatures from 10~K to 165~K in 5~K intervals (heating ramp 5~K/min).
The spectral differences between the ice deposited for the TPD experiment and the ices used for the irradiation experiments are minimal, as shown in the Appendix Figure \ref{fig:slow_fast_TPD}.

\begin{figure*}[h!]
    \sidecaption
    \includegraphics[width=12cm]{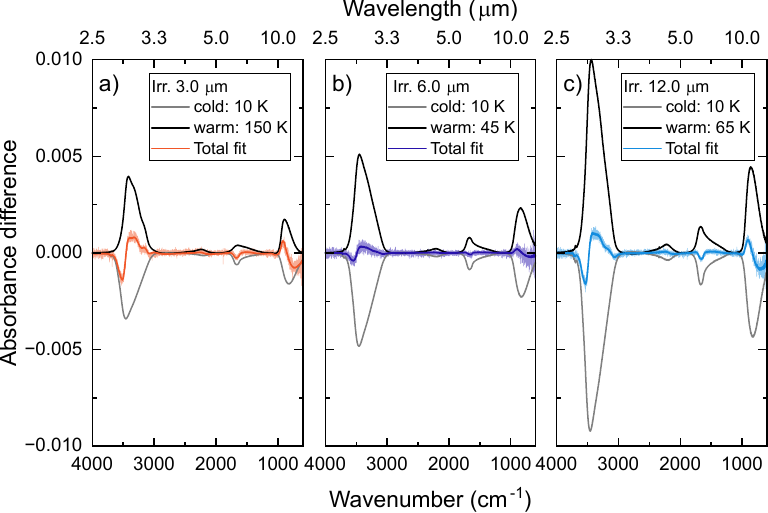}
    \caption{Linear combination fits of the three exemplary irradiations on-resonance with a) the OH-stretching vibration (3.0~$\mu$m with 47~mJ, orange), b) the bending vibration (6.0~$\mu$m with 55~mJ, blue) and c) the libration mode (12.0~$\mu$m with 134~mJ, cyan) with a cold (10~K) and a warm infrared spectrum obtained from a reference TPD experiment.
    The cold component is indicated in grey, and the warm components of the fit, of varying temperatures, are indicated in black.}
    \label{fig:hot-coldfit}
\end{figure*}

Using the TPD references, the infrared difference profiles were fitted with a linear combination of a 10~K spectrum (cold component, $\vec{s}_{\rm{c}}$), a spectrum at an elevated temperature (warm component, $\vec{s}_{\rm{w}}$), and a fixed-value baseline correction, following
\begin{equation}
    \vec{s}_d=a\vec{s}_{\rm{c}}+b\vec{s}_{\rm{w}} +c.
\end{equation}
For each irradiation difference spectrum, 32 linear-least-squares fits on the regions of the \ce{H2O} stretching and bending modes were performed using the 10~K spectrum for $\mathrm{s}_c$ and one of the warm reference spectra for $\mathrm{s}_w$ using an Octave in-house script \citep{octave}.  
The warm component was then selected based on visual comparison of 32 fits and confirmed by the $R^2$-value of the fit. 
The best fits for the three exemplary irradiations are shown in Figure \ref{fig:hot-coldfit}.
For all difference spectra, a negative cold-component contribution and a positive warm component contribution were obtained, confirming the gain of restructured warm material at the expense of the original cold material. 

Figure \ref{fig:hot-coldfit} already shows that the temperature of the selected warm component, which we can consider a measure for the degree of restructuring, varies significantly with irradiation conditions. This degree of restructuring reflects how much the local restructuring has progressed towards the ideal tetrahedral surrounding and we quantify this in terms of `restructuring temperature'. It is different from the extent of restructuring that we introduced earlier, which is measure for the amount of ice that has restructured in terms of column density. 
Where the exemplary 6.0~$\mu$m is best described with a global heating to 45~K, the 3.0~$\mu$m irradiation matches with an increase to 150~K, which is the crystallization temperature of \ce{H2O} to hexagonal ice in these conditions. 
This can be attributed to the observed structure in the restructuring profile in the OH-stretching region in the difference spectra. 
We can interpret the lower temperature of the warm component as a weaker restructuring that changes the pASW structure less.
In that sense, both irradiations at the libration mode and the bending modes seem cause restructuring to a lower degree compared to the OH-stretching vibration.

In principle, one could convert the areas of the cold and warm components to column densities by using the temperature-dependent band strengths and determine the column density of the desorbed molecules from the difference between these two. 
However, the areas of the two individual components are significantly larger than the net loss in integrated absorbance, as shown in Figure~\ref{fig:hot-coldfit}. 
Whereas the absolute column densities have values with a small relative uncertainty, the determined desorbed column density has an uncertainty of roughly the same order or higher than its value. 
Therefore, we consider these values nonsensical and refrain from using them.

In the case of most 3.0~$\mu$m irradiations, we observed that a two-component fit did not result in a satisfactory agreement, especially at long irradiation times. 
For these longer irradiations of up to 5~min, the temperature of the warm component of the best fit appeared to go down with time, while the difference spectrum did not appear to have less substructure compared to irradiations during shorter times.
Additionally, it appears counterintuitive that the degree of restructuring would decrease, and as such, previously restructured material would return to its initial disordered state.
Therefore, we constructed a three-component fit, with an additional hot component represented by a 165~K crystalline spectrum, allowing the fit to present two different populations with different degrees of restructuring. 
This three-component fitting procedure was first applied to five spectra of consecutive irradiations on the same spot to investigate the influence of irradiation time.

\begin{figure*}
    \sidecaption
    \includegraphics[width=12cm]{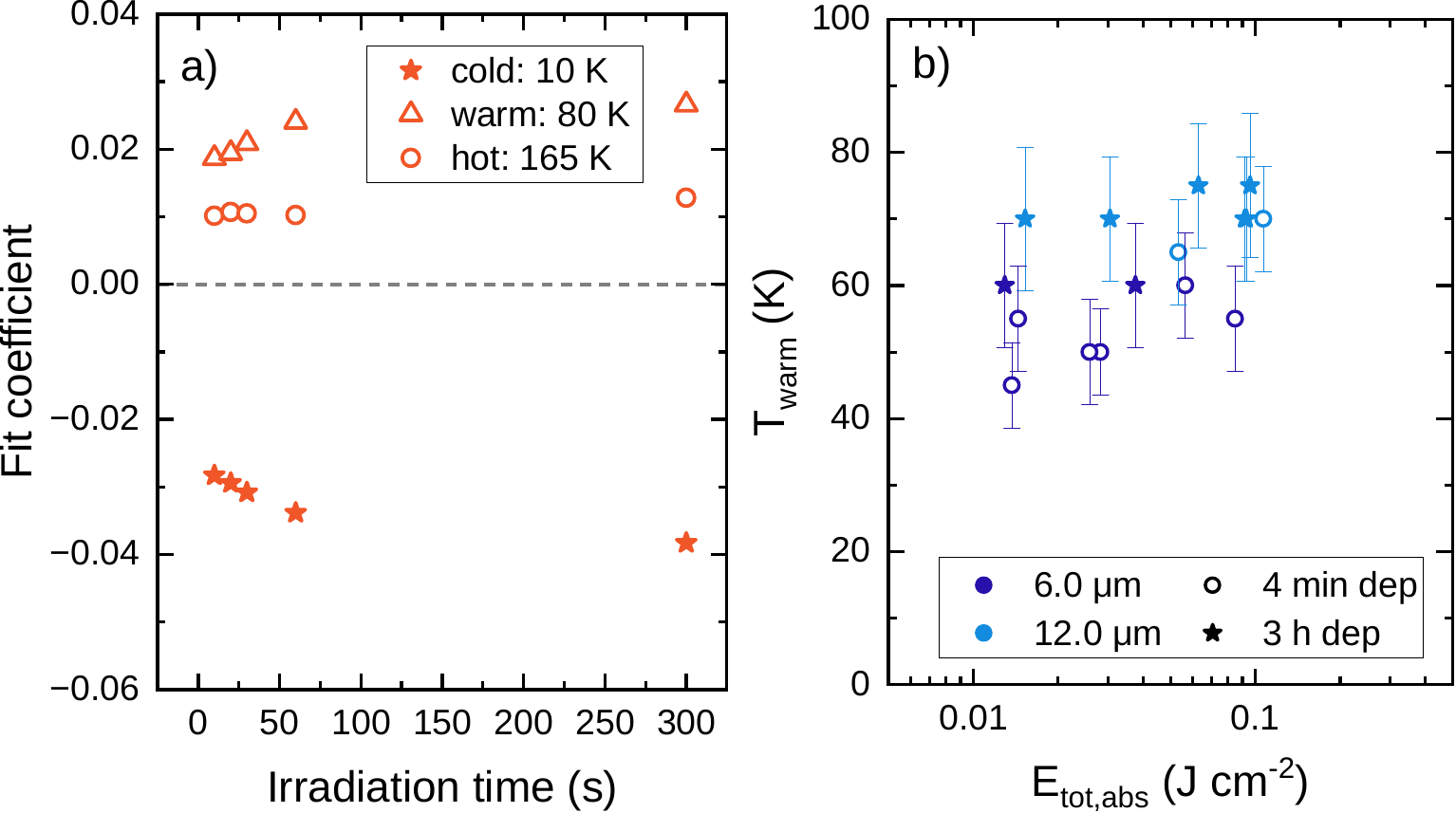}
    \caption{a) Coefficients of the contributions of the cold component (10~K, stars), the warm component (80~K, open triangles) and the hot component (165~K, circles) to the three-component fit of the infrared difference spectra of five consecutive irradiations at 3.0~$\mu$m with 48~mJ and varying irradiation time on the fast deposited ice (4 min deposition). b) Warm-component temperatures, $T_\text{warm}$, from the two-component fits to the infrared difference spectra of irradiations on-resonance with the bending vibration (6.0~$\mu$m, blue) and the libration mode (12.0~$\mu$m, cyan) on a pASW sample deposited during 3 h (stars) and a pASW sample deposited in 4 min (circles). The error bars are determined from the standard deviation of the $T_\text{warm}$ of the fits within $R^2\pm0.006$ with respect to the highest $R^2$.}    
    \label{fig:two_comp_fit}
\end{figure*}

Figure \ref{fig:two_comp_fit}a) shows the coefficients of the contributions of the amorphous cold (10~K) component, the crystalline hot (165~K) component and the warm component. 
For these irradiations, the best fit was obtained with a warm component of 80~K, below the crystallization temperature. 
The three-component fits of three of these irradiations are shown in the Appendix Figure \ref{fig:3comp_fit}.
The coefficients of the three components are related to $A_\text{change}$, but the effects of desorption and restructuring to different extents are separated in the three components.

The loss component shows very similar results compared to $A_\text{change}$: an initial strong change followed by a more gradual change at long irradiation time. 
The positive coefficients show that this change initially corresponds to very strong restructuring, locally similar to a crystalline environment. 
For longer irradiation times, the restructuring is better described by the warm component. 
The contribution of the 80~K component increases, suggesting that the additional population that is restructured -- as indicated by the increasingly negative amorphous component -- grows in size, but is not restructured as strongly as the population restructured during the first part of the irradiation.
This again hints at two different restructuring processes occurring at two different time scales, where the slower process is less strong or efficient compared to the initial restructuring during the first 5~s of the irradiation, resulting in restructuring to a lower degree.
This also explains why the two-component fit resulted in decreasing temperature for the warm component, as clearly the contribution of the 80~K component in Figure \ref{fig:two_comp_fit}a) is growing, while the crystalline contribution remains the same, such that the total difference spectra get a stronger low-temperature character at longer irradiation times.

\subsection{Trends in the restructuring and desorption} \label{sec:trends}
To further study how the restructuring and desorption depend on the irradiation conditions of the FEL and the two regimes we observe in both processes, we present our irradiation results based on two different quantities; \emph{(i)} the energy absorbed, $E_{\text{tot,abs}}$ and \emph{(ii)} the number of photons absorbed per molecule, $N_{\gamma\text{,tot,abs}}$, both estimated in the top monolayer of the ice during the total irradiation time.
Both are determined from the macropulse energy $E_\text{macro}$ that is measured during the FEL irradiations.
The details of these estimations are reported in Appendix \ref{app:Ngamma}.

Notably, the calculation of the number of absorbed photons takes into account the photon wavelength, whereas the calculation of the absorbed energy does not.
As such, $N_{\gamma\text{,tot,abs}}$ includes the fact that photons at 3.0~$\mu$m carry more energy than photons at 6.0~$\mu$m.
Additionally, we consider the absorbed quantities by correcting the incident photons or energy for the fraction absorbed, which depends on the absorption coefficient of the different vibrational modes. 
As described before, the OH-stretching vibration has a larger absorption coefficient, such that the absorbed fraction of the OH-stretching vibration is significantly larger compared to that of the bending and libration vibrations.
Naturally, we continue to apply the correction for the difference in irradiated areas across the three wavelengths.

Since we expect desorption to be a surface process, we only consider the energy adsorbed in the top monolayer. 
Depending on the mechanism, the energy absorbed in the entire multilayer ice might be more representative as is likely the case for the bulk processes of the restructuring.
Still, the absolute thickness of the ices cannot be determined sufficiently accurately from the current dataset, and as the exact mechanism fueling desorption from pASW is not yet known, we choose to avoid additional uncertainties in our analysis by considering only the energy absorbed by the top monolayer. 
Besides, since the measurements are all for ices of similar thickness, this choice should not affect the observed trends. 

Figure \ref{fig:two_comp_fit}b) shows the warm component temperatures ($T_\text{warm}$) of the two-component fits of the 6.0~$\mu$m and 12.0~$\mu$m irradiations to show a possible trend in the degree of restructuring as a function of the total absorbed energy during the irradiation, $E_{\text{tot,abs}}$.
The error bars are determined from the warm component temperatures of the fit that have an $R^2$ value within 0.006 of the best fit with the highest $R^2$.

Figure \ref{fig:two_comp_fit}b) does reveal a slight trend as a function of absorbed energy, although the $T_\text{warm}$ appears to be rather independent of the absorbed energy. 
Instead, the photon energy appears to be a more determining factor in the restructuring efficiency: irradiations at 12.0~$\mu$m generally result in higher warm-component temperatures compared to irradiations at 6.0~$\mu$m, whereas $E_\text{tot,abs}$ is rather similar.
Possibly, this is related to the vibrational movement that is triggered by the excitation of a particular mode and the influence of this mode on the hydrogen-bonding network that was found to be essential for the restructuring in pASW previously \citep{noble2020,cuppen2022}.
The libration movement stimulates a rotation of the water molecules that allows them to reorient with respect to the surrounding molecules, whereas the
motion of the bending vibration is almost perpendicular to the hydrogen bonds and can therefore not induce reorientation or restructuring.
The weaker coupling of the bending mode to the hydrogen-bonding network likely reduces the efficiency of energy dissipation through it and, therefore, may reduce the efficiency of the restructuring.
This is reflected in the lower warm temperatures obtained when fitting the difference spectra, and as such, the degree of restructuring appears to be mode dependent.

\begin{figure*}[h]
    \sidecaption
    \includegraphics[width=12cm]{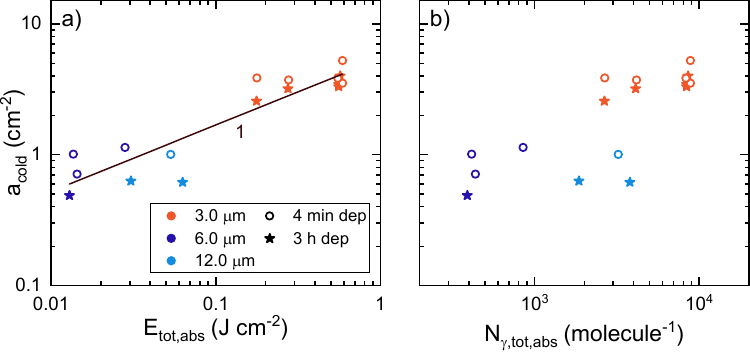}
    \caption{Absolute value of the coefficient of the cold component of the two- and three-component fits of the OH-stretching region in the infrared difference spectra corrected for the irradiated area as a function of a) the total energy absorbed during the full irradiation, $E_{\text{tot,abs}}$ and b) the total number of photons absorbed per molecule during the full irradiation, $N_{\gamma\text{,tot,abs}}$. The irradiations are performed at the OH-stretching vibration (3.0~$\mu$m, orange), the bending vibration (6.0~$\mu$m, blue) and the libration mode (12.0~$\mu$m, cyan) on a pASW sample deposited during 3 h (solid stars) and a pASW sample deposited in 4 min (open circles). Only irradiations of 30~s are shown. The solid line shows a linear fit as a function of macropulse energy.}
    \label{fig:E_Ac}
\end{figure*}

Figures \ref{fig:E_Ac}a) and \ref{fig:E_Ac}b) show the absolute value of the coefficient of the cold component of the two-component and three-component fits to the infrared difference spectra $a_\text{cold}$ of the irradiations as a function of $E_{\text{tot,abs}}$ and $N_{\gamma\text{,tot,abs}}$, respectively. 
The $a_\text{cold}$ values are a measure for the extent of the restructuring, like the $A_\text{change}$ values.
We chose to analyze the cold component instead of the warm component of the fit, as this value is independent of the different spectral shapes associated with the various warm-component temperatures.
Additionally, $a_\text{cold}$ is similar when obtained from a two-component or three-component fit and therefore allows for the combination of the results for both fits in this analysis.
In case of the irradiations on the bending vibration (6.0~$\mu$m, blue) and the libration mode (12.0~$\mu$m, cyan), where hardly any desorption is observed, $a_\text{cold}$ obtained from the two-component fits reflects the amount of material that has restructured during the irradiation.
For the irradiations on the OH-stretching vibration (3.0~$\mu$m, orange), $a_\text{cold}$ from the three-component fit could be influenced by the desorption of material, which we cannot disentangle. 
Still, assuming that the desorption is a surface process connected to a limited population, we postulate that this contribution will be small compared to the restructuring.

Figure \ref{fig:E_Ac} shows the $a_\text{cold}$ values for the irradiations of equal duration (30~s) only.
This is to exclude the time-dependent effect observed for the irradiations on the OH-stretching vibration from the analysis of the dependence on the total absorbed energy.
A clear linear trend on the log-log scale is observed in Figure \ref{fig:E_Ac}a), as shown by the solid line labelled 1.
The slope of this line is reported in Table \ref{tab:power}.
Such a trend on the log-log scale suggests exponential behavior of the restructuring as a function of the absorbed energy.

Although line 1 is only fitted to the irradiations of 30~s time duration, Appendix Figure \ref{fig:E_asc_app} shows that the slope of a linear fit of all 42 irradiations performed with varying irradiation times is within the error margin on the slope of the fit.
This suggests that irradiations on all vibrational modes result in the same dependence on the absorbed energy.
Still, this does not mean that the process of restructuring is fully vibrational mode independent.
Although $a_\text{cold}$, a measure for the extent of the restructuring, shows no clear vibrational mode dependence, $T_\text{warm}$ in Figure \ref{fig:two_comp_fit}b) reveals that the degree of restructuring is slightly mode dependent.
As such, the degree of restructuring is likely vibrational mode dependent, while the extent of the restructuring is not.

The difference between Figures \ref{fig:E_Ac}a) and \ref{fig:E_Ac}b) lies in the photon energy and hence the trends within symbols of the same color remain the same, whereas the groups of colored symbols change relative to each other. 
For instance, irradiations at 12.0~$\mu$m have similar $N_{\gamma\text{,tot,abs}}$ values compared to irradiations at 3.0~$\mu$m, but significantly different $E_{\text{tot,abs}}$ values.
In general, the absorbed energy and absorbed photons per molecule show differences of less than an order of magnitude in the distribution of the $a_\text{cold}$ values.

Although the 6.0~$\mu$m and 12.0~$\mu$m irradiations are included in the fit in Figure \ref{fig:E_Ac}a) and they seem to show a similar trend, we cannot directly compare trends in the libration and bending modes with irradiations of the OH-stretching vibration because the range in $E_{\text{tot,abs}}$ and $N_{\gamma\text{,tot,abs}}$ is limited for the libration and bending modes.
This is inherent to the experiments due to several factors.
Firstly, the OH-stretching is an intense band, and as such absorbs a larger fraction of the incoming photons compared to the less intense libration and bending modes.
As this is a physical property of the system, it can only be overcome by increasing the irradiation energy for the weaker modes or decreasing the energy for the OH-stretching vibration.

Decreasing the irradiation energy for irradiations on-resonance with the stretching mode complicates the analysis because the sensitivity for restructuring and desorption is reduced due to the smaller irradiation area compared to longer wavelength irradiations. 
For example, the lowest intensity irradiations (13 dB attenuation) at 3.0~$\mu$m result in difference spectra with a restructuring signal that is hardly observed above the noise.
The $a_\text{cold}$ values for these irradiations are therefore not determined, as the low signal-to-noise did not result in a sensible two-component or three-component fit, whereas in the same regime of absorbed energy, the changes for irradiation at 6.0~$\mu$m and 12.0~$\mu$m are clear and $a_\text{cold}$ can be obtained reliably. 

Increasing the irradiation energy for the bending and libration modes also has its limitations, as the current reported number of photons and absorbed energy are already at the maximum power of the FEL at these wavelengths.
If one would irradiate the libration mode with the same number of absorbed photons as the OH-stretching vibration irradiation with the highest desorption signal (0.009 absorbed photons per molecule per micropulse), the FEL should supply a macropulse power of $\sim325$~mJ.
This is impossible for the FELIX FEL-2. 
In this way, it cannot be tested whether, for higher numbers of absorbed photons, the restructuring and desorption from irradiations at the bending and libration modes show the same trends as the irradiations at the OH-stretching vibration or whether the libration mode may result in more efficient restructuring.

Interestingly the 12.0~$\mu$m irradiations show similar values for $a_\text{cold}$ at similar absorbed energies compared to 6.0~$\mu$m irradiations.
Combining this with the observations from Figure \ref{fig:two_comp_fit}b), irradiation at the bending mode results in weaker restructuring, as indicated by $T_\text{warm}$ compared to irradiation of the libration mode, while the extent of the irradiation is comparable as parametrized by $a_\text{cold}$.
This highlights that the libration mode is more efficient in restructuring compared to the bending mode, indicated by the higher degree of restructuring, but is equally efficient in dissipating the vibrational energy, as indicated by the extent of the restructuring. 

So far, we mainly focused on the restructuring occurring during irradiation, yet we also observed desorption during the irradiation experiments.
To properly analyze the mass spectrometry data, we use a Python script to extract the maximum intensity of each desorption spike occurring at a 5~Hz frequency, corresponding to the FEL macropulse frequency.
Additionally, we extract a baseline signal from the same 5~Hz pattern, shifted by half the time between FEL macropulses, relative to the extracted desorption spikes.
We use the baseline signal as a baseline correction on the desorption signal.

\begin{figure*}
    \sidecaption
    \includegraphics[width=12cm]{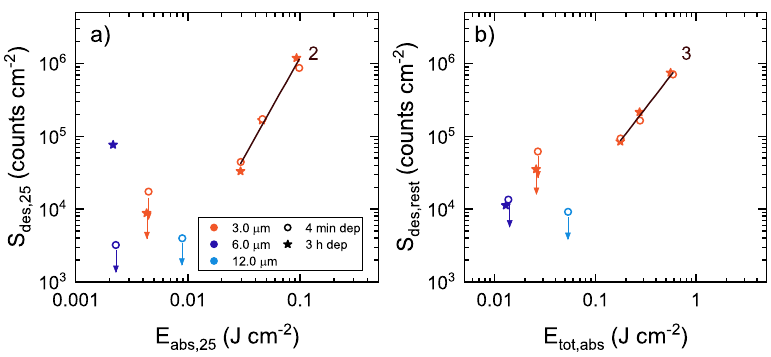}
    \caption{Desorption signals measured for $m/z$~18 (\ce{H2O}) by the mass spectrometer in MID mode as a function of a) the energy absorbed during the first 25 macropulses, $E_{\text{abs,}25}$ and b) the energy absorbed during the full irradiation, $E_{\text{tot,abs}}$.  Irradiations are performed during 30~s at the OH-stretching vibration (3.0~$\mu$m, orange), the bending vibration (6.0~$\mu$m, blue) and the libration mode (12.0~$\mu$m, cyan) on a pASW sample deposited during 3 h (solid stars) and a pASW sample deposited in 4 min (open circles). MID traces with no clear desorption or only a single desorption peak are treated as upper limits, indicated with downwards arrows. The solid lines indicate the linear trends on the log-log scale reported in Table \ref{tab:power}.}
    \label{fig:Ng_S}
\end{figure*}

Based on the two regimes observed in the mass spectrometry measurement, we define two quantities to represent these different regimes: \emph{(i)} the sum of the intensity of the first 25 desorption spikes $S_{\text{des},25}$, representing the region of the exponential decrease in the desorption intensity and \emph{(ii)} the sum of the remaining desorption spikes recorded during the rest of the irradiation $S_{\text{des,rest}}$, representing the region of constant desorption.
Figure \ref{fig:Ng_S} shows a) $S_{\text{des},25}$ as a function of $E_{\text{abs,25}}$, the absorbed energy during the first 25 macropulses and b) $S_{\text{des,rest}}$ as a function of the total absorbed energy $E_{\text{tot,abs}}$.
$E_{\text{abs,25}}$ and $E_{\text{tot,abs}}$ are both determined for the absorption of energy in the first monolayer of the ice.
Both the desorption during the first 25 macropulses and the rest of the irradiations show a linear trend on the logarithmic scales. 
In Figure \ref{fig:Ng_S}b), this trend seems to continue to the lower absorbed energies of the 6.0~$\mu$m and 12.0~$\mu$m irradiations, although these values are all upper limits except one.

From these linear trends on the log-log scale indicated in Figures \ref{fig:E_Ac} and \ref{fig:Ng_S} for the restructuring and desorption, respectively, we can extract the exponential dependencies of the processes, using that for $y=Ax^n$ on a log-log scale, $n$ can be extracted from the slope of a linear fit.
The slopes and Pearson's r values for these linear fits are reported in Table \ref{tab:power}.
From these power dependencies, we find that restructuring takes place with the square root of the absorbed energy, whereas the desorption appears to be a higher order process. 
It should be noted here that we cannot distinguish the two regimes in $a_\textbf{cold}$ as we can for the desorption measurements, and the order of the process could describe a combination of fast and slow processes.
Still, as $a_\textbf{cold}$ describes the extent of the restructuring, where we observe that already 70~\% of the restructuring has taken place within 10~s, it is likely that $a_\textbf{cold}$ for the 30~s irradiations in Figure \ref{fig:E_Ac}a) mainly characterizes the first time regime of fast processes.

The square root order of the dependence of the extent of the restructuring on the absorbed energy means that the process becomes less efficient with an increasing number of photons and more absorbed energy. 
As mentioned above, we postulate that a limited number of molecules is available for restructuring, while a single excitation can lead to restructuring in a somewhat extended area. 
During continued irradiation and exposure to more photons, an increasing number of molecules have already restructured. 
As a result, the photon energy must be dissipated over a larger area to affect molecules that are still available for restructuring.
This matches well with the observed increase in the extent of the restructuring, while the degree of the restructuring decreases.
The result is the combination observed from the three-component fits of a larger population that has been heated to above the crystallization temperature and a smaller population that is heated to a lesser extent. 

For the desorption, the process appears to be of a higher order. 
Although these are sometimes related to multiphoton processes, this cannot be the case in our experiments.
An estimate of the highest number of photons absorbed per molecule per micropulse in the first monolayer of the ice for our strongest irradiation is only 0.01.
After this micropulse, the energy is dissipated readily before the next micropulse will interact with the ice, excluding the chances of multiphoton processes in our experiments \citep{cuppen2022,schrauwen2024,schrauwen2025CH4}.
Instead, we expect the higher-order process to result from long-distance interactions between excited molecules.
Such long-distance interactions are likely in a hydrogen-bonding network, as seen in \ce{H2O}.
In this way, the desorption requires multiple photons, but not by the direct absorption of multiple photons by one molecule or its closest neighbors, but by the interaction of various excited areas over longer distances.

\begin{table}[t]
    \centering
    \caption{Results of the linear fits of Figures \ref{fig:E_Ac} and \ref{fig:Ng_S}.}
    \begin{tabular}{l l r r}
        \hline line\tablefootmark{a} & variable & slope & Pearson's r \\ \hline \hline
        1 & $a_\text{cold}$ & 0.51 (0.06) & 0.919 \\
        2 & $S_\text{des,25}$ & 2.75 (0.20) & 0.989 \\
        3 & $S_\text{des,rest}$ & 1.80 (0.10) & 0.994 \\ \hline
    \end{tabular}
    \label{tab:power}
    \tablefoot{
      Slopes and Pearson's $r$ of the linear fits of $\log(X)$ with $X=a_\text{cold}$, $S_\text{des,25}$ or $S_\text{des,rest}$ as a function of $\ln\left(E_\text{macro}\right)$, the natural logarithm of the macropulse energy. \\
      \tablefoottext{a}{The line numbers refer to the numbered lines in Figures \ref{fig:E_Ac} and \ref{fig:Ng_S}.}
}
\end{table}
Both the restructuring and desorption seem to influence relatively large populations, fuelled by the efficient energy dissipation through the hydrogen-bonding network of \ce{H2O}.
The first few FEL macropulses likely influence the material directly inside the highest energy segment of the irradiation spot, especially considering that the FEL has a Gaussian intensity profile.
This corresponds to the first regime, where we observe the strongest restructuring and desorption during the first 5~s or 25 macropulses.
The second regime, with continued weak desorption and the growth of a restructured `warm' population, then arizes from the efficient energy dissipation through the hydrogen-bonding network to an increasingly larger area.
Considering that the irradiated spot is relatively small compared to the size of the ice, the continued restructuring and desorption in the second regime are only limited by the reach of the energy dissipation in the hydrogen-bonding network.
As such, the second regime seems to address a much larger population compared to the first regime, but due to the long-distance energy dissipation that is required, the efficiency of the processes in this regime is less. 

Due to the significantly different slopes determined for the desorption and restructuring processes, it appears unlikely that the restructuring and desorption are linked processes or address the same population.
If our infrared irradiation resulted in on-resonance heating, we could expect both restructuring and desorption to be linked to the heating of the ice system through similar processes with similar orders.
Moreover, one would naively expect that crystallization should occur prior to desorption.
Our experiments cannot probe the order of the desorption and restructuring processes, as the shortest irradiation time measuring restructuring lasted for 10~s.
Yet, a consecutive nature of restructuring and desorption would link these processes, which is refuted by their significantly different orders.

Additionally, the slight vibrational mode dependence observed suggests that the specific vibration influences the restructuring process, especially when comparing the two-component fits for the 6.0~$\mu$m and 12.0~$\mu$m irradiations. 
As a consequence, the degree of the restructuring is mode dependent and therefore not linked to the absorbed energy and likely not a thermal process.
Then, even though the fitting routines discussed here use spectra of heated ices as a reference and indicate that the restructuring resembles a thermal effect, the origin of the restructuring is likely non-thermal.
Moreover, thermal processes induced by the excitation of a vibrational mode would drive the system rapidly to equipartition.
Since we still observed differences in the degree of restructuring depending on the different vibrational modes excited, this likely does not happen.
Therefore, we conclude that the desorption and restructuring that we observe are not necessarily thermal processes.

\section{Conclusion and astrophysical Implications}
Irradiations on all three vibrational modes reveal the characteristic down-up profile connected to a restructuring of the pASW ice, as observed before \citep{noble2020,cuppen2022}.
At the same time, we clearly observed the desorption of \ce{H2O} from pure porous amorphous water ice upon the irradiation of the OH-stretching vibration that was previously only reported for crystalline \ce{H2O} and a CO-\ce{H2O} system \citep{krasnapoler1998,slumstrup2025}.
Little to no desorption of \ce{H2O} is observed for irradiation of the bending and libration modes.

Analysis of the down-up profile upon restructuring shows that 
that irradiations of the bending mode and libration mode result in weaker restructuring in terms of local structural change, resembling heating to below 100~K, compared to the OH-stretching vibration, which is best fitted with heating above the crystallization temperature of 155~K.
Although we cannot directly compare the results of the OH-stretch irradiations with the irradiations of the bending and libration vibrations because of the differences in FEL intensity and the size of the irradiated area, these results suggest that the resulting structure depends on the excitation mode, whereas the extent of this restructuring, the number of molecules affected, appears to be independent of the excited vibrational mode and mainly depends on the absorbed energy. The libration mode restructures more efficiently in terms of structural for the same amount of absorbed energy compared to the bending vibration.

The desorption and restructuring happen in parallel, but the trends as a function of absorbed energy or absorbed photons suggest very different dependencies.
The restructuring decreases in efficiency with increasing absorbed energy, whereas the desorption appears to be a higher-order process: up to a third-order process for the first 25~macropulses of the irradiation. 
Considering that at maximum 0.01 photons are absorbed per molecule, an actual multiphoton process is unlikely, and we expect long-range interactions between excited areas to be responsible for the higher-order desorption process.
Because of these different orders of the restructuring and desorption processes, it is unlikely that both processes have the same origin, ruling out that both the restructuring and desorption are the result of thermal heating of the ice.

Both the restructuring and desorption reveal two different regimes:
\begin{enumerate}
    \item a regime during the first $\sim$25 macropulses with the strongest restructuring and desorption of the irradiation 
    \item a regime during the rest of the irradiation with significantly weaker effects, but continued restructuring and desorption
\end{enumerate}
In this first regime, the irradiation results in $\sim$70~\% of the extent of restructuring of the longest irradiation and strong desorption that exponentially decays with every macropulse.
The exponential decay suggests that the desorption is limited by the available population, which is likely made up of loosely bound molecules and not the complete desorption of the ice as observed by \cite{krasnapoler1998} for crystalline \ce{H2O} at 110~K.
The weak power dependence also suggests that it is rather limited by the number of molecules available of restructuring and not by the number of photons. This would mean that we can compare experimental results to fluences available in dense molecular clouds. The first restructuring effects are observed for an absorbed energy of 0.01 J cm$^{-2}$ in the experiments. The same fluence is reached after 360 years in space. This is fast compared to the time it takes to build up an appreciable ice layer of roughly 10$^6$ years and hence we expect IR processing to help in the annealing of the ice. Since the process occurs during ice build-up, the ice layer will be constantly replenished, and saturation, as observed in the experiments, is less likely to occur.
Moreover, infrared-induced photoprocessing proceeds through the excitation of vibrational nodes and since the dissipation of excess reaction energy from chemical reactions  can also result in vibrational excitation, the same mechanism may facilitate reactive restructuring.

After the population available for direct restructuring and desorption is depleted during the first regime of restructuring, the continued processes observed in the second regime rely on the efficient dissipation of energy through the hydrogen-bonding network of \ce{H2O}.
In this way, an area larger than the initial irradiated area leads to continued desorption and restructuring at a reduced rate. 
This could be related to the Gaussian intensity profile of the FEL beam, where the center of the irradiation spot generally receives a higher irradiation intensity.
Hovever, it is also possible that the efficient energy dissipation through the hydrogen-bonding network is responsible for the spreading of the restructuring.

The experiments also showed two regimes for infrared photodesorption and we believe they have the same underlying origin: depletion of the material during the first regime leading to a different mechanism in the second regime. Previous infrared irradiation studies of interstellar ice analogs have not decoupled the two different regimes observed here, and the analysis of desorption from mass spectrometry data has focused on the first regime while analysis from infrared spectra was on the combined effect \citep{noble2020,cuppen2022,schrauwen2024,ingman2023,slumstrup2025}.
Although the reason for the regime change might be the same, the different orders for desorption and restructuring hint that the underlying mechanisms in these regimes might be different.
The high power dependence for the desorption in the first regime is less likely to be relevant for interstellar ices in the ISM, where the infrared radiation field is significantly weaker compared to that of the FEL and multi-photon processes, or processes relying on a considerable photon influx cannot occur. The second regime is likely more relevant where a constant low desorption rate was observed that we could quantify as on the order of $10^{15}$ \ce{H2O} molecules J$^{-1}$ (see Appendix~\ref{app:desorption}). For the CO desorption upon 3.0~$\mu$m excitation of the underlying p-ASW explored in \cite{slumstrup2025}, the same calculation yields a desorption rate on the order of $10^{16}$ CO molecules J$^{-1}$. The order of magnitude more efficient desorption confirms that the indirect photodesorption of CO is very efficient. For the current study, we cannot exclude similar indirect processes, and if this is the case, the binding energy of the desorbing molecule appears to be the determining factor. 

The experimental desorption rates can be related to ISM conditions. Here we work with ISM fluxes of absorbed photons/energy by multiplying the broadband IR flux with the IR absorption cross section. This treatment effectively separates out the bending, stretching and libration wavelength ranges, which has also been covered experimentally.
The flux of absorbed IR energy in the ISM is on the order of $10^{-12}$ J s$^{-1}$ cm$^{-2}$, meaning that our experimental results correspond to an ISM desorption rate of $10^{3}$~\ce{H2O} molecules~s$^{-1}$~cm$^{-2}$.
For UV-induced \ce{H2O} desorption, the desorption rate has been  experimentally found to be on the order of $10^{-3}$ molecules per incident UV photon \citep{Bulak2023, Fillion2022}. Applying the cosmic ray-induced UV photon flux in dense clouds on the order of $10^{4}$~photons~s$^{-1}$~cm$^{-2}$ yields a UV-induced ISM desorption rate of ca. 10 molecules~s$^{-1}$~cm$^{-2}$. 
This is two orders of magnitude lower than the IR-induced ISM desorption rate estimated in this work, indicating that the IR-induced desorption processes could be the dominant desorption route in dense, interstellar clouds. The main caveat is in the uncertainty of the photon order and the exact desorption mechanism and rate. These are difficult to determine for the second, continuous slow-desorption regime, which is the most relevant for ISM conditions, and around the detection limit of desorption in our experiments. 

Future studies should focus more on this second regime by investigating longer irradiation times, lower-power irradiations and repeated irradiations to gather the statistics necessary to determine restructuring and desorption rates that are outside of the scope of this current paper.
At the same time, a better understanding of the origin of the high-order behavior of the first desorption regime would indicate if the desorption mechanism in this regime could be relevant in ISM conditions as well.
These studies require a better overlap of the FEL and the infrared spectrometer for irradiations at 3.0~$\mu$m to increase sensitivity to the small changes induced by prolonged irradiation, as well as irradiations with lower FEL intensities to extend the range of absorbed energies and achieve a better match with irradiations at 6.0~$\mu$m and 12.0~$\mu$m.
Then, by further studying the restructuring and desorption of \ce{H2O} ice, the influence of the omnipresent infrared radiation field in the ISM can be better described and modeled.

\begin{acknowledgements}
\textbf{JGMS} acknowledges this publication as part of the project ``HFML-FELIX: a Dutch Center of Excellence for Science under Extreme Conditions'' (with project number 184.035.011) of the research program ``Nationale Roadmap Grootschalige Wetenschappelijke Infastructuur'' which is (partly) financed by the Dutch Research Council (NWO).
The main components of the LISA experimental apparatus were purchased using funding obtained from the Royal Society through grants UF130409, RGF/EA/180306, and URF/R/191018.
\textbf{SI}, \textbf{LS}, and \textbf{JDT} thank the Danish National Research Foundation through the Center of Excellence ``InterCat'' (Grant agreement no.: DNRF150) for financial support.
All authors thank the FELIX technical staff for maintenance and operation of the FELIX FELs and for support in maintenance of the LISA setup.
\end{acknowledgements}

\begin{appendix}
\section{Spectral comparison of irradiated ices with TPD reference} \label{app:spec_comp}
\begin{figure}[h]
    \centering
    \includegraphics[width=\linewidth]{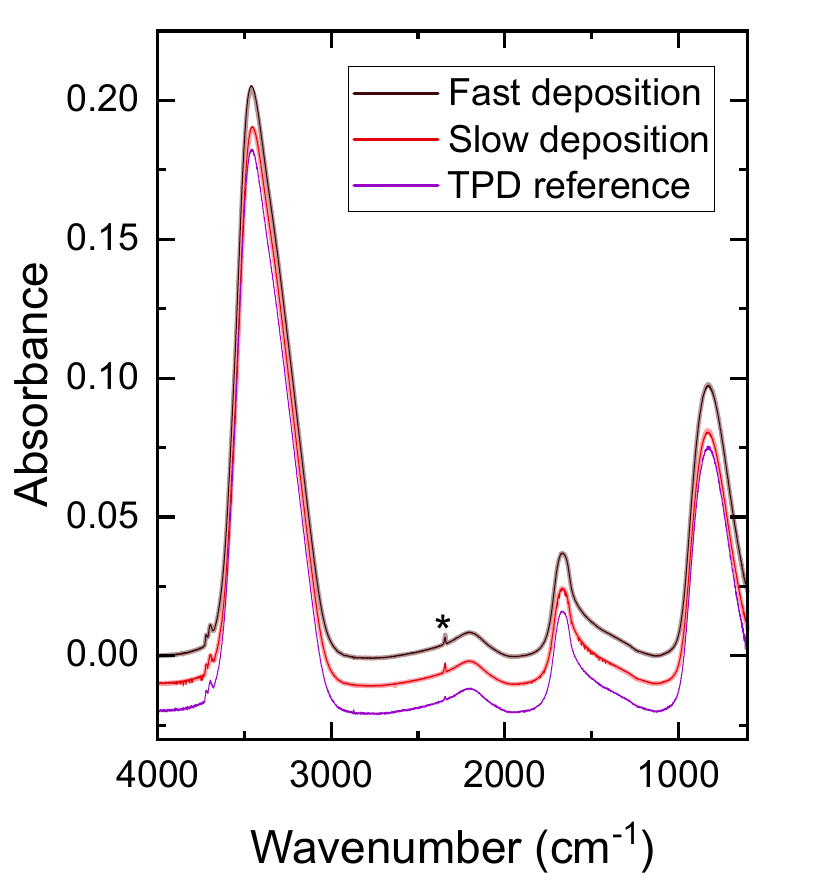}
    \caption{RAIR spectra of two pASW ices deposited during three hours (red, light red), two pASW ices deposited in 4 min (brown, light brown) and a pASW sample deposited in 4 min for the TPD reference measurement (purple). Of the duplicate pASW depositions, one of the ices is shown in a thicker trace behind the other, as to highlight their similarity. The spectra are offset for clarity, and the asterisk indicates a small contamination of \ce{CO2}.}
    \label{fig:slow_fast_TPD}
\end{figure}

Figure \ref{fig:slow_fast_TPD} shows the RAIR spectra of the pASW ices deposited for this paper. 
The TPD reference is deposited during 4 minutes as well.
A small contamination of \ce{CO2} can be observed in the spectra, but this did not influence the measurements as no changes are observed in this mode and no desorption of \ce{CO2} was detected by the mass spectrometer.
The pASW ices deposited during 4 minutes (purple and black curves) have slightly higher  OH stretch and libration features compared to the spectrum of the slow deposition (red curve). This effect is less pronounced for the bending feature. These differences might hint at a different thickness or at a slightly different structure of the ices, although overall the spectral shape for the three ices are very similar.

\section{MID and MCS comparison} \label{app:MCS}
\begin{figure}
    \centering
    \includegraphics[width=\linewidth]{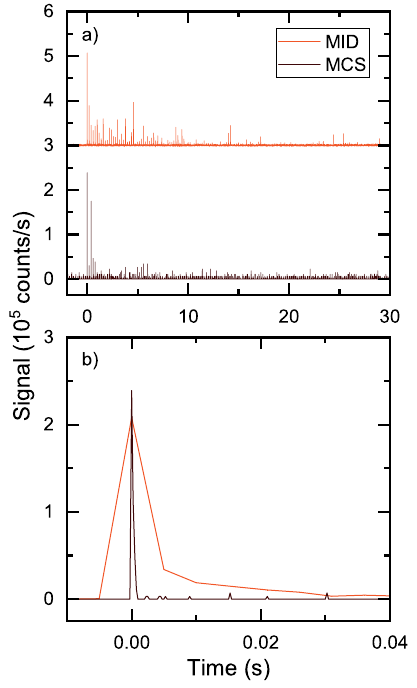}
    \caption{Comparison of the \textit{m/z}~18 signal obtained during irradiation using the unsynchronized MID mode with a 5~ms time resolution (orange) and using the synchronized MCS mode (brown) with a 0.25~ms resolution for two separate irradiations at 5~Hz on the OH-stretching vibration of pASW. Both traces have been baselined.
    b) shows a zoom in on the first macropulse.}
    \label{fig:MCSvsMID}
\end{figure}

Some time after the experiments performed for this paper were conducted, a new mode for the mass spectrometry measurements was installed.
Where the conventional MID mode of the mass spectrometer is not synchronized with the FEL and presents a 0.2~ms downtime per data point, the new multichannel scaler (MCS) based mode is synchronized to the FEL trigger.
In MCS mode, the mass spectrometer measures $m/z$~18 with a time resolution of 0.25~ms triggered by the FEL and without the downtime observed in MID mode, resulting in an imaging of the desorption spike as a result of the macropulse with significantly higher resolution, as shown in Figure \ref{fig:MCSvsMID}b).
The MCS mode has not yet been fully optimized to measure desorption events, but it will present a promising tool for recording infrared photodesorption in future research.

The traces in Figure \ref{fig:MCSvsMID} were recorded for two separate samples studied in different beamshifts for different projects, with a several-month interval, and therefore do not depict the same desorption events.
The MID data correspond to a 30~s irradiation with 48~mJ at 5~Hz on a pASW ice deposited for 4~min~40~s at $5\cdot10^{-6}$ mbar, while the MCS data were for a 1~min irradiation with 37~mJ at 5~Hz on a pASW ice deposited for 4~min~20~s at $5\cdot10^{-6}$ mbar.
Although aimed at different projects, the ices are spectroscopically very similar and yet less desorption is expected for the pASW ice with a lower $E_\text{macro}$ of the MCS measurements.
Clearly, the MID mode has a lower time resolution, but the lack of synchronization and the uncontrolled downtime of the measurements do not seem to affect the mass spectrometry measurements significantly.
The two regimes observed in the MID data are also present in the MCS data. 

\section{Three-component fits}
\begin{figure*}
    \sidecaption
    \includegraphics[width=12cm]{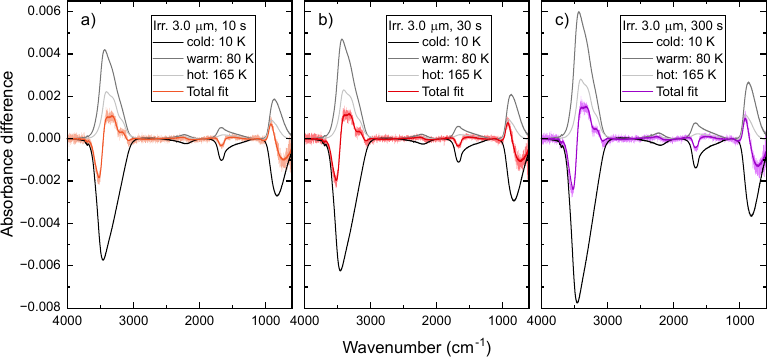}
    \caption{Linear combination fits consisting of three spectral components of three irradiations on-resonance with the OH-stretching vibration at 3.0~$\mu$m with similar irradiation intensities but different irradiation times of a) 10~s, b) 30~s and c) 300~s. The fitted difference spectrum is included in the background in a lighter shade of the color of the fit in the foreground. The cold component of the fit was fixed with a 10~K infrared spectrum obtained from a reference TPD experiment (black) and the highest temperature component was fixed at 165~K (light grey).
    The best fits were obtained with a warm component of 80~K as indicated in grey.}
    \label{fig:3comp_fit}
\end{figure*}

Figure \ref{fig:3comp_fit} shows the three-component fit for three irradiations at 3.0~$\mu$m with similar irradiation intensities, but different irradiation times of 10, 30 and 300 s.
The three-component fit consists of a 10~K cold component, a 165~K hot or crystalline component and a warm component of varying temperatures. 
For these three irradiations, the best fit based on the $R^2$ and visual inspection involved a warm component of 80~K.

\section{Estimation of the absorbed energy and the number of absorbed photons} \label{app:Ngamma}
During the operation of the FEL, the beam intensity is monitored continuously.
The intensity is reported as energy per macropulse, $E_\text{macro}$, in units of Joules.
From this value, we can determine the total amount of energy that was absorbed by the ice during the irradiation within the irradiated spot, using the following
\begin{equation}\label{eq:Eabs}
    E_\text{tot,abs}=\frac{4\sin(45^\circ)ftE_\text{macro}}{\pi d^2}\left(1-\exp(-\alpha l)\right)
\end{equation}
with $f$ the FEL macropulse frequency, $t$ the irradiation time and $\frac{\pi d^2}{4\sin(45^\circ)}$ the area of the spot on the substrate irradiated by the FEL beam with a diameter $d$ and an angle with the substrate of $45^\circ$.
The factor $\left(1-\exp(-\alpha l)\right)$ corrects for the fact that, depending on the absorption coefficient $\alpha$ of the specific vibrational mode and the thickness of the absorbing layer $l$, only a fraction of the impinging energy is absorbed.
In this case, we only consider the amount of energy absorbed by the first monolayer ($l=3$~\AA{}), as the deeper monolayers will systematically absorb less, such that our estimation reflects the maximum amount of energy that can be absorbed in a monolayer.
Naturally, if the restructuring and desorption would be bulk processes, an absorbed fraction based on the full ice thickness would be more representative.
Yet, from our data, we cannot determine if the process is related to the bulk or surface, and we have no reliable way to determine the thickness of the ice.
Then, to prevent any additional uncertainties in the representation of the data we only consider the absorption in one monolayer. 

The absorption coefficient $\alpha$ is vibrational mode dependent and is also related to the frequency overlap of the FEL with the vibrational mode.
To include these factors in the absorbed fraction, we determined $\alpha$ using
\begin{equation}
    \alpha=2\pi k\tilde{\nu}
\end{equation}
with $k$ the extinction coefficient and $\tilde{\nu}$ the specific wavelength that $\alpha$ is determined for.
Based on the $k$ values from \citep{rocha2023} and the approximate width of the FEL in our experiments, we find values of approximately $1.6\cdot10^4$, $0.1\cdot10^4$ and $0.4\cdot10^4$ for irradiation at 3.0~$\mu$m, 6.0~$\mu$m and 12.0~$\mu$m, respectively.

In the case of the FEL, the diameter of the beam is also wavelength dependent.
Using a heat-sensitive camera (Spiricon Pyrocam IV), the size of the FEL beam containing 95~\% of the light was determined to be 0.73~mm at 3.0~$\mu$m, 1.46 mm at 6.0~$\mu$m, 1.43~mm at 8.0~$\mu$m and 2.11~mm at 15~$\mu$m.
The value of 1.90~mm for 12.0~$\mu$m was obtained by extrapolation of the 8.0~$\mu$m and 15.0~$\mu$m values, as described in \citet{slumstrup2025}.

Besides the absorbed energy, we also consider the number of absorbed photons per molecule, $N_{\gamma,\text{abs}}$, using
\begin{equation}\label{eq:Ngammatot}
    N_{\gamma,\text{tot,abs}}=\frac{E_\text{tot,abs}\lambda}{ hc\rho_\text{surf}}=  \frac{4\sin(45^\circ)ft\lambda E_\text{macro}}{\pi hcd^2\rho_\text{surf}}\left(1-\exp(-\alpha l)\right)
\end{equation}
with $\lambda$ the irradiation wavelength, $h$ the Planck constant, $c$ the speed of light and $\rho_\text{surf}$ the average surface density of an ice ($10^{15}$~cm$^{-2}$).
We use the same vibrational mode-dependent factor $\left(1-\exp(-\alpha l)\right)$ to correct for the number of photons that are not absorbed in the first monolayer.

Previous experiments have shown that the vibrational excitation in a pASW ice is dissipated readily within 1~ns, which is less than the time between two micropulses \citep{cuppen2022}.
Then, to estimate whether a single molecule can absorb multiple photons in a single micropulse, we can use 
\begin{equation}\label{eq:Ngamma}
    N_{\gamma,\text{abs}}=\frac{4\sin(45^\circ)\lambda E_\text{macro}}{\pi hcd^2\rho_\text{surf}N_\text{micro}}\left(1-\exp(-\alpha l)\right)
\end{equation}
with $N_\text{micro}$ the number of micropulses in a macropulse, which is 6000 for a 6~$\mu$s macropulse and a micropulse frequency of 1~GHz.
For the strongest irradiation performed at 2.99~$\mu$m with 48~mJ, we estimate 0.01 photons absorbed per molecule using Equation \ref{eq:Ngamma}.
As a result, the absorption of multiple photons by a single molecule or/and its close neighbors is impossible and multi-photon effects are unlikely.

\section{Trend in absorbed energy for all irradiations}
\begin{figure*}
    \sidecaption
    \includegraphics[width=12cm]{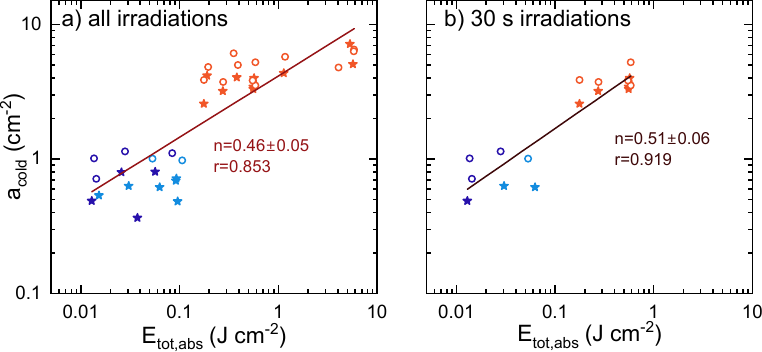}
    \caption{Coefficient of the cold component of the two- and three-component fits of the OH-stretching region in the infrared difference spectra corrected for the irradiated area as a function of the total energy absorbed during the full irradiation, $E_{\text{tot,abs}}$ for a) all irradiations and b) a selection of irradiations with equal duration of 30~s. The irradiations are performed at the OH-stretching vibration (3.0~$\mu$m, orange), the bending vibration (6.0~$\mu$m, blue) and the libration mode (12.0~$\mu$m, cyan) on a pASW sample deposited during 3 h (solid stars) and a pASW sample deposited in 4 min (open circles). The solid lines show the linear fit as a function of macropulse energy, including the slope $n$ and Pearson's r.}\label{fig:E_asc_app}
\end{figure*}

Figure \ref{fig:E_asc_app} shows the coefficients of the cold component of the fits as a function of the total absorbed energy during the irradiation $E_{\text{tot,abs}}$ for a) all 42 irradiations and b) the selection of irradiations with equal duration (30~s) shown in the main text.
The solid lines present the linear fits on the log-log scale for a) all irradiations and b) the selection of irradiation with equal duration.
The slopes of these linear fits are consistent within their error margins, indicating that the order of the process is largely unaffected by the duration of the irradiation.

\section{Quantification of the water desorption rate}
\label{app:desorption}
Water desorption is measured by the mass spectrometer but to quantify the desorption rate, the mass spectrometry count needs to be calibrated. For CO desorption from a non-porous ASW layer, the signal was high enough to measure desorption by both mass spectrometry and IR spectroscopy \citep{slumstrup2025}. We used a 3~$\mu$m irradiation of CO-on-np-ASW (13 layers of CO, 11mJ), measured in the same setup, to calibrate the desorption count in the mass spectrometer in MCS mode. Comparing the decrease in CO column density from IR to the total CO counts in the mass spectrometer during irradiation results in a factor of $2.8 \times 10^{9}$ CO molecules/cm$^2$ per mass spectrometer count/cm$^2$, considering the irradiation spot size.
Accounting for the difference in sensitivity of the mass spectrometer for CO and \ce{H2O} \citep{Martin-Domenech2015}, we obtain an estimated factor of $2.2 \times 10^{9}$ \ce{H2O} molecules/cm$^2$ per mass spectrometer count/cm$^2$.

With this factor, the desorption signals presented in Figure~\ref{fig:Ng_S} can be converted to estimates for the number of \ce{H2O} molecules desorbed. As an example, the data points in Figure~\ref{fig:Ng_S}b) described by line 3 represent approximately $0.2-1.6\times 10^{15}$ \ce{H2O} molecules cm$^{-2}$ for $E_\text{tot,abs} = 0.18-0.58$~J cm$^{-2}$, resulting in a desorption rate in the range of $1.1-2.8 \times 10^{15}$ \ce{H2O} molecules J$^{-1}$.

\end{appendix}

\end{document}